\documentclass[aps,prfluids,onecolumn,superscriptaddress,amsmath,amssymb]{revtex4-2}
\usepackage[utf8]{inputenc}   
\usepackage[T1]{fontenc}
\usepackage{lmodern}
\usepackage[english]{babel}
\usepackage{graphicx}
\usepackage{bm}
\usepackage{overpic}
\usepackage{geometry}
\usepackage{color}
\usepackage{amsmath,amsfonts,amssymb}
\usepackage{subfigure}
\usepackage{rotating}
\usepackage[subfigure]{tocloft}
\usepackage{hyperref} 

\DeclareMathOperator*{\argmax}{arg\,max}

\usepackage[]{xcolor}

\begin{document}

\title{Lacrymaria-like dynamics: compression and length change in active slender bodies}
\author{Clément Moreau}
 \affiliation{
	Nantes Université, École Centrale Nantes, IMT Atlantique, CNRS, LS2N, UMR 6004, F-44000 Nantes, France}
\author{Yukinori Nishigami}
 \affiliation{
	Research Center of Mathematics for Social Creativity, Research Institute for Electronic Science, Hokkaido University, Kita-ku, Sapporo 001-0020, Japan}
\affiliation{Graduate School of Life Science, Hokkaido University, Kita-ku, Sapporo 060-0810, Japan}
\author{Johann Herault}
 \affiliation{
	Nantes Université, École Centrale Nantes, IMT Atlantique, CNRS, LS2N, UMR 6004, F-44000 Nantes, France}
\author{Kenta Ishimoto}
 \affiliation{
	Department of Mathematics, Kyoto University, Kyoto 606-8502, Japan}
\date{\today}

\begin{abstract}
Length change is central to the behaviour of many active slender bodies, yet most elastohydrodynamic models of microscopic locomotion treat filaments as effectively inextensible. Here we introduce a compressible $N$-link formulation for planar active filaments, in which both bending angles and segment lengths are dynamical variables coupled through local hydrodynamic drag, modelled using resistive force theory. We first use the model to test whether passive extensibility can improve bending-driven swimming and crawling. Across the parameter ranges studied, passive compressibility usually reduces locomotion speed, indicating that axial compliance alone is not a robust mechanism for propulsion enhancement. We then prescribe active rest-length waves and show that compression can either enhance, suppress, or redirect locomotion depending on its phase, wavelength, and frequency relative to bending. Finally, motivated by the extensible neck dynamics of \textit{Lacrymaria olor}, we apply the method to clamped and free filaments driven by oscillatory follower forces. These examples show that elastic compression-bending coupling can reorganise swimming, exploration, and locomotion in strongly deforming active filaments.
\end{abstract}

\maketitle

\section{Introduction}

Undulatory locomotion is widespread in biology, from fish swimming and snake crawling to the motion of nematodes and unicellular microorganisms. At cellular scales, the mechanics of flexible slender bodies has been central to understanding how microswimmers generate propulsion in low-Reynolds-number fluids, where viscous forces dominate inertia. Classical examples include the whip-like motions of cilia and flagella \cite{taylor1951analysis, gray1955propulsion, lauga2020fluid, wan2019ciliate}.

In most elastohydrodynamic models, cilia and flagella are treated as inextensible elastic rods, because axial compression and elongation are usually small compared with bending deformation. Filament extensibility has sometimes been introduced for numerical regularisation \cite{simons2015fully,ishimoto2018elastohydrodynamical,lagrone2019elastohydrodynamics}, but its mechanical role in bending-driven microscopic locomotion remains comparatively underexplored.

There are, however, important biological and artificial systems in which length change is not a small correction. A prominent example is \textit{Lacrymaria} \citep{yanase2018neck, flaum2024curved}, a unicellular organism whose neck can extend to several times the body length \cite{qin2024dynamic}. Large compressive or extensile deformations also occur in other unicellular organisms such as \textit{Euglena} \citep{arroyo2012reverse} and \textit{Stentor} \citep{huang1973contractile, echigoya2026geometrical}. The spasmoneme of \textit{Vorticella}, although structurally distinct from cilia and flagella, can undergo rapid calcium-triggered contraction \citep{ryu2016vorticella}. Extensibility is also relevant in multicellular small organisms such as nematodes \citep{meneely2019working}, whose swimming occurs in low-Reynolds-number conditions \cite{montenegro2016flow}, and in artificial microrobots based on hydrogels \citep{nikolov2015self, sharan2021fundamental, tan2024shape} or self-assembled droplets \citep{cholakova2021rechargeable, lisicki2024twist}. On a macroscopic scale, compression plays an equally important role in some modes of animal locomotion, such as rectilinear motion of snakes \cite{newman2018crawling}. 

Active compression has been widely studied in crawling systems such as gastropods, nematodes, and earthworms \cite{desimone2012crawling, tanaka2012mechanics, paoletti2014proprioceptive, kanazawa2024locomotion}, and compressibility has been implemented in models of passive filaments \cite{garg2023slender}. By contrast, few studies have addressed undulatory microswimming with extensible filaments \cite{pak2011extensibility}. Recent works have revisited locomotion generated by coupled bending and compression in a compressive scallop model \cite{gidoni2024gait,ishimoto2025bending}, showing that one bending mode and one compression mode can generate net displacement at zero Reynolds number. Perturbation analyses further indicate that compression can enhance manoeuvrability when coupled to bending \cite{ishimoto2025bending}. These studies prescribe the shape kinematics, however, leaving open how axial compliance affects locomotion when the filament shape emerges from elastohydrodynamic dynamics.

Here, we develop a compressible $N$-link elastohydrodynamic model for active slender bodies. We first consider coarse-grained $N$-link descriptions, in which an elastic filament is described by slender rigid segments connected by Hookean elastic hinges \cite{moreau2018asymptotic, gazzola2018forward, walker2020efficient, alouges2025n}, and can be seen as a coarse-grained representation of an inextensible Kirchhoff rod. Hydrodynamic drag is modelled as a local anisotropic operator according to standard resistive force theory \cite{gray1955propulsion}. The model we introduce extends this class to compressible unshearable Reissner rods by treating the segment lengths as dynamical variables alongside the bending angles, allowing large axial deformation while retaining a finite-dimensional formulation.

We use the model to analyse passive and active compressibility in bending-driven locomotion, considering both swimming and crawling by varying the drag anisotropy ratio in resistive-force theory \cite{keaveny2017predicting, ishimoto2025robust}. In the parameter ranges studied here, passive compressibility usually reduces propulsion, whereas active compression has a more nuanced effect and can enhance, suppress, or redirect locomotion depending on its coupling to bending.

We then apply the same framework to large-amplitude \textit{Lacrymaria}-like neck dynamics. \textit{Lacrymaria} extends and contracts its neck using activity concentrated near the mouth at the distal tip. During repeated extension and contraction, the neck direction can be reorganised by buckling instabilities, producing exploratory behaviour during foraging \cite{coyle2019coupled}. Krishnamurthy and Prakash \cite{krishnamurthy2023emergent} proposed a minimal follower-force model for this behaviour, in which a flexible filament is driven by a time-dependent tangential force at its tip. Related follower-force models have also been used for microtubule dynamics \cite{de2017spontaneous, ling2018instability, man2019morphological,clarke2024bifurcations}. The model of Ref.~\cite{krishnamurthy2023emergent} reproduced periodic, aperiodic, and chaotic behaviours reminiscent of \textit{Lacrymaria}, but focused primarily on bending instabilities, with the role of neck compressibility remaining less resolved. Here, we investigate the effect of large compressibility on the follower-force model, and report qualitative changes at high activity. Our results suggest that compressibility may enhance exploration and locomotion behaviours.

The paper is organised as follows. Section~\ref{sec:model} introduces the compressible $N$-link formulation. Section~\ref{sec:locomotion} examines passive and active compression in bending-driven swimming and crawling. Section~\ref{sec:follower} applies the method to clamped and free follower-force filaments inspired by \textit{Lacrymaria}. Section~\ref{sec:discussion} discusses the implications and limitations of the model.

\section{Coarse-grained model}
\label{sec:model}

\begin{figure}
    \centering
    \includegraphics[width=0.66\linewidth]{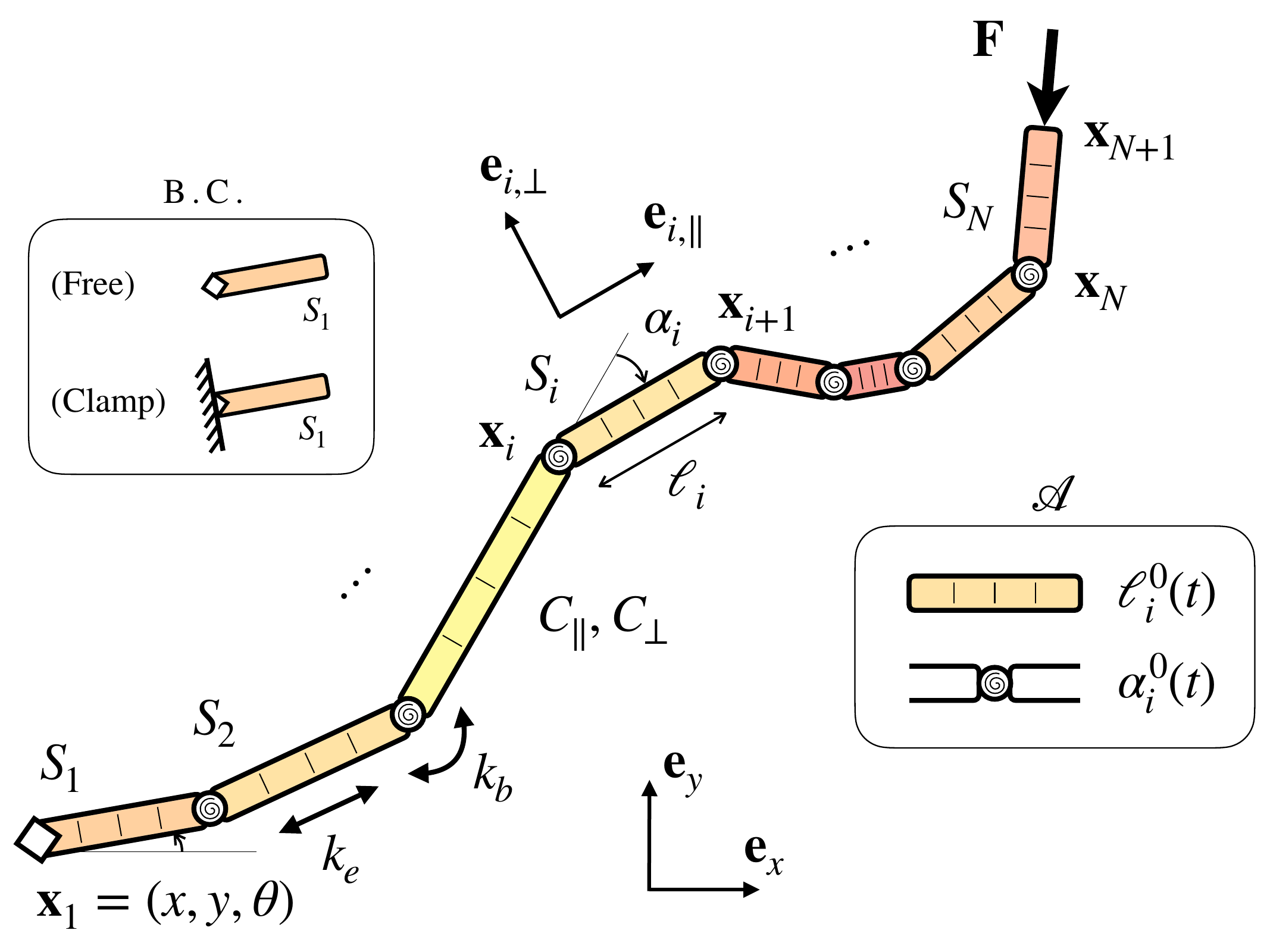}
    \caption{Diagram of the compressive $N$-link model}
    \label{fig:setup}
\end{figure}

\subsection{Kinematics}

To describe a compressible filament, we use an $N$-link, or $N$-segment, coarse-grained representation \cite{moreau2018asymptotic}. The planar filament is discretised into $N$ slender straight segments $S_i$, $i=1,\dots,N$; see Figure~\ref{fig:setup}. The instantaneous length of segment $S_i$ is denoted by $\ell_i(t)$, so that changes in the variables $\ell_i$ represent local extension or compression. The filament has total length $L$ and we write $\ell = L/N$. 
Successive segments are connected by junctions $J_j$, $j=1,\dots,N-1$. The relative angle between $S_j$ and $S_{j+1}$ is denoted by $\alpha_j(t)$ and represents the discrete bending deformation.

The filament motion is assumed to remain in the plane spanned by $(\mathbf{e}_x,\mathbf{e}_y)$, with $\mathbf{e}_z = \mathbf{e}_x \times \mathbf{e}_y$. The endpoint of the first segment has position $\mathbf{x}=(x,y)$, and the first segment makes an angle $\theta$ with $\mathbf{e}_x$. For each segment $S_i$, we denote by $\mathbf{x}_i=(x_i,y_i)$ its proximal endpoint and by $\theta_i$ its absolute orientation, so that $\mathbf{x}_1=\mathbf{x}$ and $\theta_1=\theta$. The distal endpoint of the filament is $\mathbf{x}_{N+1}$. We also introduce the segment basis $(\mathbf{e}_{i,\parallel},\mathbf{e}_{i,\bot})$, where $\mathbf{e}_{i,\parallel}$ is directed along segment $S_i$ and $\mathbf{e}_{i,\bot} = \mathbf{e}_z \times \mathbf{e}_{i,\parallel}$.

The filament can be described by $2N+2$ absolute endpoint coordinates,
$\mathbf{X}_{\mathrm{abs}} = (x_1,\dots,x_{N+1},y_1,\dots,y_{N+1})$. 
Equivalently, we use the relative coordinates
$\mathbf{X}_{\mathrm{rel}} = (x,y,\theta,\alpha_1,\dots,\alpha_{N-1},\ell_1,\dots,\ell_N)$.
The two descriptions are related by
\begin{equation}
    \alpha_j = \theta_{j+1} - \theta_j, \quad
    \ell_i  = \| \mathbf{x}_{i+1} - \mathbf{x}_i \| ,
    \quad j=1,\dots,N-1,\quad i=1,\dots,N,
    \label{eq:x_abs_to_rel}
\end{equation}
and, conversely,
\begin{equation}
    \theta_i = \theta + \sum_{j=1}^{i-1} \alpha_j, \quad
    \mathbf{x}_{i+1} = \mathbf{x}_{i} + \ell_i \mathbf{e}_{i,\parallel}.
    \label{eq:x_rel_to_abs}
\end{equation}
For linear and angular velocities, we differentiate Eqs.~\eqref{eq:x_abs_to_rel} and \eqref{eq:x_rel_to_abs} with respect to time and obtain:
\begin{equation}
    \dot{\alpha}_j = \dot{\theta}_{j+1} - \dot{\theta}_{j}, \quad
    \dot{\mathbf{x}}_{i+1} = \dot{\mathbf{x}}_{i}
    + \ell_i \dot{\theta}_i \mathbf{e}_{i,\bot}
    + \dot{\ell}_i \mathbf{e}_{i,\parallel}.
    \label{eq: velocities}
\end{equation}

\subsection{Dynamics}

\subsubsection{Hydrodynamics}
\label{sec:hydro}

The filament is immersed in a zero-Reynolds-number fluid, and we approximate the hydrodynamic drag on each segment using resistive-force theory \cite{gray1955propulsion}. For a material point $\mathbf{x}_c=(x_c,y_c)$ on segment $S_i$, we denote by $\mathbf{u}_{\parallel}$ and $\mathbf{u}_{\bot}$ the components of its velocity along $\mathbf{e}_{i,\parallel}$ and $\mathbf{e}_{i,\bot}$. The local drag force density $\mathbf{f}_c$ is then
\begin{equation}
    \mathbf{f}_c = - (C_{\parallel} \mathbf{u}_{\parallel} + C_{\bot} \mathbf{u}_{\bot}) = - (C_{\parallel} (\dot{\mathbf{x}}_c \cdot \mathbf{e}_{i,\parallel}) \mathbf{e}_{i,\parallel} + C_{\bot} (\dot{\mathbf{x}}_c \cdot \mathbf{e}_{i,\bot}) \mathbf{e}_{i,\bot}),
    \label{eq: rft}
\end{equation}
where the drag coefficients $C_{\parallel}$ and $C_{\bot}$ depend on the body geometry and surrounding medium. For an infinitely slender body in a Newtonian fluid, $C_{\parallel}/C_{\bot}=1/2$. Integrating over segment $S_i$ gives the hydrodynamic force $\mathbf{F}_i$ and hydrodynamic torque, projected along $\mathbf{e}_z$, about a junction point $\mathbf{x}_j$:
\begin{equation}
    \mathbf{F}_i = \int_{S_i} \mathbf{f}_c (\mathbf{x}) \mathrm{d}\mathbf{x}, \quad T_i^{\mathbf{x}_j} = \mathbf{e}_z \cdot \int_{S_i} (\mathbf{x} - \mathbf{x}_j) \times \mathbf{f}_c (\mathbf{x})\mathrm{d} \mathbf{x}.
    \label{eq:hydro_force_torque}
\end{equation}
To close the equations for length changes, we also compute the hydrodynamic force integrated over the distal half-segment $S_{i,1/2}=[\ell_i/2,\ell_i]$ of segment $S_i$,
\begin{equation}
    \mathbf{F}_i^+ = \int_{S_{i,1/2}} \mathbf{f}_c (\mathbf{x}) \mathrm{d}\mathbf{x}.
\end{equation}
Exploiting Eqs.~\eqref{eq: velocities} and \eqref{eq: rft} to compute the integrals in \eqref{eq:hydro_force_torque} yields
\begin{align}
    \mathbf{F}_i &
    = - \ell_i \left[ C_{\parallel} 
    \left( \dot{\mathbf{x}}_i \cdot \mathbf{e}_{i,\parallel} 
    + \frac{\dot{\ell}_i}{2} \right) \mathbf{e}_{i,\parallel} 
    + C_{\bot} \left( \dot{\mathbf{x}}_i \cdot \mathbf{e}_{i,\bot} 
    + \frac{\ell_i \dot{\theta}_i}{2} \right) \mathbf{e}_{i,\bot} \right], \label{eq:F_hyd} \\
    \mathbf{F}_i^{+} & = - \ell_i \left[ C_{\parallel} 
    \left( \frac{\dot{\mathbf{x}}_i \cdot \mathbf{e}_{i,\parallel}}{2} 
    + \frac{3\dot{\ell}_i}{8} \right) \mathbf{e}_{i,\parallel} 
    + C_{\bot} \left( 
    \frac{\dot{\mathbf{x}}_i \cdot \mathbf{e}_{i,\bot}}{2} 
    + \frac{3 \ell_i \dot{\theta}_i}{8} \right) 
    \mathbf{e}_{i,\bot} \right], \label{eq:F_hyd_half} \\    
    T^{\mathbf{x}_j}_i & =  \ell_i \left[ 
    C_{\parallel}\, d_{ji,\bot} 
    \left( \dot{\mathbf{x}}_i \cdot \mathbf{e}_{i,\parallel} 
    + \frac{\dot{\ell}_i}{2} \right) 
    - C_{\bot} \left( 
    \left( d_{ji,\parallel} + \frac{\ell_i}{2} \right) 
    \left( \dot{\mathbf{x}}_i \cdot \mathbf{e}_{i,\bot} \right) 
    + \dot{\theta}_i \ell_i \left( \frac{d_{ji,\parallel}}{2} 
    + \frac{\ell_i}{3} \right) 
    \right) \right].
    \label{eq:T_hyd}
\end{align}
In Eq. \eqref{eq:T_hyd}, we have introduced the displacement vector $\mathbf{d}_{ji} := \mathbf{x}_i - \mathbf{x}_j$, with components $d_{ji,\parallel} = \mathbf{d}_{ji} \cdot \mathbf{e}_{i,\parallel}$ and $d_{ji,\bot} = \mathbf{d}_{ji} \cdot \mathbf{e}_{i,\bot}$.

\subsubsection{Elasticity}

We adopt a linear elastic model for both bending and stretching deformations. At each junction $J_j$, a Hookean restoring torque opposes deviations from a rest angle $\alpha_{0,j}$:
\begin{equation}
    T_j^{\mathrm{el}} = - \frac{k_b}{\ell} \left( \alpha_j - \alpha_{0,j} \right),
    \label{eq:T_el}
\end{equation}
where $k_b > 0$ is the continuous bending stiffness. Similarly, each segment $S_i$ exerts a Hookean restoring force on its neighbours opposing deviations from some rest length $\ell_{0,i}$:
\begin{equation}
    \mathbf{F}_i^{\mathrm{el}} = - \frac{k_e}{\ell} \left( \ell_i - \ell_{0,i} \right) 
    \mathbf{e}_{i,\parallel},
    \label{eq:F_el}
\end{equation}
where $k_e > 0$ is the stretching stiffness. We denote by $L$ the total filament length at rest, $L = \sum_{i=1}^{N} \ell_{0,i}$.

\subsubsection{Activity}
\label{sec:activity}

The filament may be set into motion by forces and torques of various origin: bend and shear from molecular motors in living flagella and cilia, or reaction to external fields such as magnetic or acoustic actuation in the case of micro-robots.  
Two types of model are considered here to broadly encapsulate these effects.

\paragraph{Internal or external actuation.} 
One may prescribe forces $\mathbf{F}_{\mathrm{act}}(\mathbf{x},t)$ and torques $T_{\mathrm{act}}(\mathbf{x},t)$ acting directly on segments and junctions, accounting for instance for molecular motors, ciliary pumping, or external fields. These active forces and torques enter the global force and torque balances, which will be introduced in the next section. In the following, we will consider in particular the \emph{follower force} 
\begin{equation}
    \mathbf{F}_{\mathrm{act}} = F_{\mathrm{act}} \mathbf{e}_{N,\parallel},
    \label{eq: follower force}
\end{equation}
applied at the tip $\mathbf{x}_{N+1}$ of the filament and always directed along the last segment. The follower force model has been widely used  in micromotility studies to effectively describe the power-and-recovery stroke observed in cilia \cite{de2017spontaneous,ling2018instability,bayly2016steady,clarke2024bifurcations}.

\paragraph{Preferred geometric state.}
Alternatively, activity can be encoded in time-varying rest shapes by taking $\alpha_{0,i}(t)$ and $\ell_{0,i}(t)$ as prescribed functions of time. In the literature, this viewpoint is often termed preferred deformation, or spontaneous deformation, and is widely used as a descriptor of the periodic internal efforts driving the deformation of active microswimmers \cite{olson2011coupling,elgeti2015physics}. The elastic restoring terms \eqref{eq:T_el}--\eqref{eq:F_el} then drive the filament toward the desired configuration at each instant. For a basic travelling wave, we consider
\begin{align}
        \alpha_{0,j}(t) & = \tau_b \sin\left(\omega_b t - \nu_b j\right), \label{eq:alpha0_wave} \\ 
        \ell_{0,i}(t) & = \ell_{0,i} + \tau_e \sin (\omega_e t - \nu_e i + \phi),
    \label{eq:ell0_wave}
\end{align}
corresponding to waves with pulsation $\omega_b$, $\omega_e$, wavenumber $\nu_b$, $\nu_e$ and amplitude $\tau_b$, $\tau_e$, respectively for bending and extensional deformations, 
with an additional phase shift $\phi$ between bending and stretching waves.

\subsubsection{Boundary conditions}

To close the system, we impose boundary conditions at both endpoints of the filament. Three cases are considered for the proximal endpoint $\mathbf{x}_1$. In the \emph{free} case, no constraint is imposed on $\mathbf{x}$ or $\theta$, and the endpoint moves freely under the filament dynamics. In the \emph{pinned} case, the position is prescribed, $\mathbf{x}(t) = \mathbf{x}_0(t)$, while the angle $\theta$ remains free. In the \emph{clamped} case, both position and orientation are prescribed, $\mathbf{x}(t) = \mathbf{x}_0(t)$ and $\theta(t) = \theta_0(t)$. In all cases, the other endpoint $\mathbf{x}_{N+1}$ is free unless an external force or follower force is applied there, as described in Section~\ref{sec:activity}.

\subsubsection{Nondimensional parameters}

The behaviour of the passive filament is characterised by three nondimensional parameters defined as follows:
\begin{equation}
    \mathrm{Sp}=L\left(\frac{C_{\bot}\omega_b}{k_b}\right)^{1/4}, \quad \mu=\frac{k_eL^2}{k_b}, \quad \gamma=\frac{C_{\parallel}}{C_{\bot}}.
\end{equation}
The parameter $\mathrm{Sp}$ characterises the ratio between viscous forces and elastic forces, and is historically called ``sperm number'' in biofluids studies \cite{gadelha2013counterbend}. Biological flagella and microscopic slender bodies typically span values from approximately $\mathrm{Sp} \approx 2$, corresponding to relatively rigid filaments, to $\mathrm{Sp} \approx 10$, corresponding to more flexible filaments \cite{eloy2012kinematics}. 

Axial elasticity is controlled by the nondimensional ratio $\mu$. Larger values correspond to less extensible filaments, and the inextensible limit is recovered as the axial stiffness tends to infinity.

Finally, we have introduced the drag anisotropy ratio $\gamma$, which is equal to $1/2$ for a slender filament in Stokes flow \cite{gray1955propulsion}. Other phenomenological settings, such as crawling models or granular resistive-force theory, can involve smaller values; for example, we use $\gamma=1/70$ for \textit{C. elegans}-like crawling on agar \cite{keaveny2017predicting}.
All activity parameters appearing in Eqs.~\eqref{eq: follower force}--\eqref{eq:ell0_wave} are nondimensional from this point onward, and we keep the same symbols for readability.

\subsection{Balance equations}

To solve the filament dynamics, we write hydrodynamic, elastic, and active balances over a hierarchy of subsystems. Specifically, we impose: (a) the force balance on the whole filament, projected onto 
    $\mathbf{e}_x$ and $\mathbf{e}_y$, yielding 2 scalar equations;
(b) the torque balance on $\{S_1, \dots, S_N\}$ with respect to 
    $\mathbf{x}$, yielding one scalar equation;
(c) $N-1$ torque balances on $\{S_i, \dots, S_N\}$ with respect to 
    $\mathbf{x}_i$, for $i = 2, \dots, N$, accounting for bending elasticity. This recovers the classical set of equations for an inextensible $N$-link swimmer \cite{alouges2025n,moreau2018asymptotic}. Here, however, $N$ supplementary equations are required for the segment lengths. We obtain them by introducing subsystems that begin at a half-segment, and impose (d) $N$ force balances on $\{S_i/2, \dots, S_N\}$ projected onto 
    $\mathbf{e}_{i,\parallel}$, for $i = 1, \dots, N$, where $S_i/2$ denotes the upper half 
    $[\ell_i/2, \ell_i]$ of segment $S_i$. 

This yields $2N+2$ equations in total, matching the dimension of $\dot{\mathbf{X}}_{\mathrm{rel}}$. Assembling these equations leads to the linear system
\begin{equation}
    \mathsf{M} \dot{\mathbf{X}}_{\mathrm{rel}} :=
    \begin{pmatrix}
    \mathbf{M}_{tt} & \mathbf{M}_{tr} & \mathbf{M}_{tb} & \mathbf{M}_{te} \\
    \mathbf{M}_{rt} & \mathbf{M}_{rr} & \mathbf{M}_{rb} & \mathbf{M}_{re} \\
    \mathbf{M}_{bt} & \mathbf{M}_{br} & \mathbf{M}_{bb} & \mathbf{M}_{be} \\
    \mathbf{M}_{et} & \mathbf{M}_{er} & \mathbf{M}_{eb} & \mathbf{M}_{ee}
    \end{pmatrix} 
    \begin{pmatrix} \dot{\mathbf{x}} \\ \dot{\theta} \\ \dot{\bm{\alpha}} \\ 
    \dot{\bm{\ell}} \end{pmatrix} 
    = 
    \begin{pmatrix}
        \mathbf{F}_{\mathrm{act,ext}} \\ T_{\mathrm{act,ext}} \\ 
        \mathbf{T}_b + \mathbf{T}_{\mathrm{act,int}} \\ 
        \mathbf{F}_e + \mathbf{F}_{\mathrm{act,int}}
    \end{pmatrix},
    \label{eq:linear_system}
\end{equation}
where $\mathbf{T}_b$ and $\mathbf{F}_e$ collect the bending and stretching elastic terms from \eqref{eq:T_el} and \eqref{eq:F_el}, respectively. The active terms $\mathbf{F}_{\mathrm{act,ext}}$, $T_{\mathrm{act,ext}}$, $\mathbf{T}_{\mathrm{act,int}}$, and $\mathbf{F}_{\mathrm{act,int}}$ collect the external and internal actuation contributions described in Section~\ref{sec:activity}. 
The matrix $\mathsf{M}$ encodes the hydrodynamic drag terms computed in Section~\ref{sec:hydro}, with a block structure reflecting the coupling between translational, rotational, bending, and extensional degrees of freedom, denoted by the indices $t,r,b,e$. Its assembly is detailed in Appendix~\ref{app:matrix}. Once $\mathsf{M}$ is assembled, Eq.~\eqref{eq:linear_system} is solved for the instantaneous velocity vector and integrated in time.

\subsection{Numerical implementation}

In the simulations below, the matrix $\mathsf{M}$ is assembled at each time step from the instantaneous filament configuration and the resulting linear system is solved for $\dot{\mathbf{X}}_{\mathrm{rel}}$. 
The system \eqref{eq:linear_system} can be integrated with standard ODE solvers. For relatively large values of $N$ and high deformations, the matrix $\mathsf{M}$ can become nearly singular and require a very small time step; this can be improved with specialised solvers for stiff systems such as MATLAB's \texttt{ode15s}. Numerical stability is also improved by using the segment rest length $\Delta L=L/N$ as the length unit instead of the total length $L$. In the code, the nondimensional bending-to-viscosity ratio is therefore
\[
    \mathrm{Sp}_{\Delta}
    =
    \Delta L\left(\frac{C_{\bot}\omega_b}{k_b}\right)^{1/4}
    =
    \frac{\mathrm{Sp}}{N},
\]
and the hydrodynamic matrix is multiplied by $\mathrm{Sp}_{\Delta}^4=(\mathrm{Sp}/N)^4$. Code and data are available at \url{https://github.com/Clementmoreau/compressive-elastohydrodynamics}.

\section{Impact of compression on flagellar locomotion}
\label{sec:locomotion}

Many microorganisms and cells use flexible slender appendages called flagella to move through their surroundings. Understanding how flagellar deformation and activity generate motion in the fluid has been a long-standing problem in microhydrodynamics, dating back to the seminal work of Taylor \cite{taylor1951analysis}. In this section, we use our compressible $N$-segment model to explore the effect of compressibility on locomotion abilities of slender filaments. 

\subsection{Passive compression}

\begin{figure}
    \centering
    \includegraphics[width=0.9\linewidth]{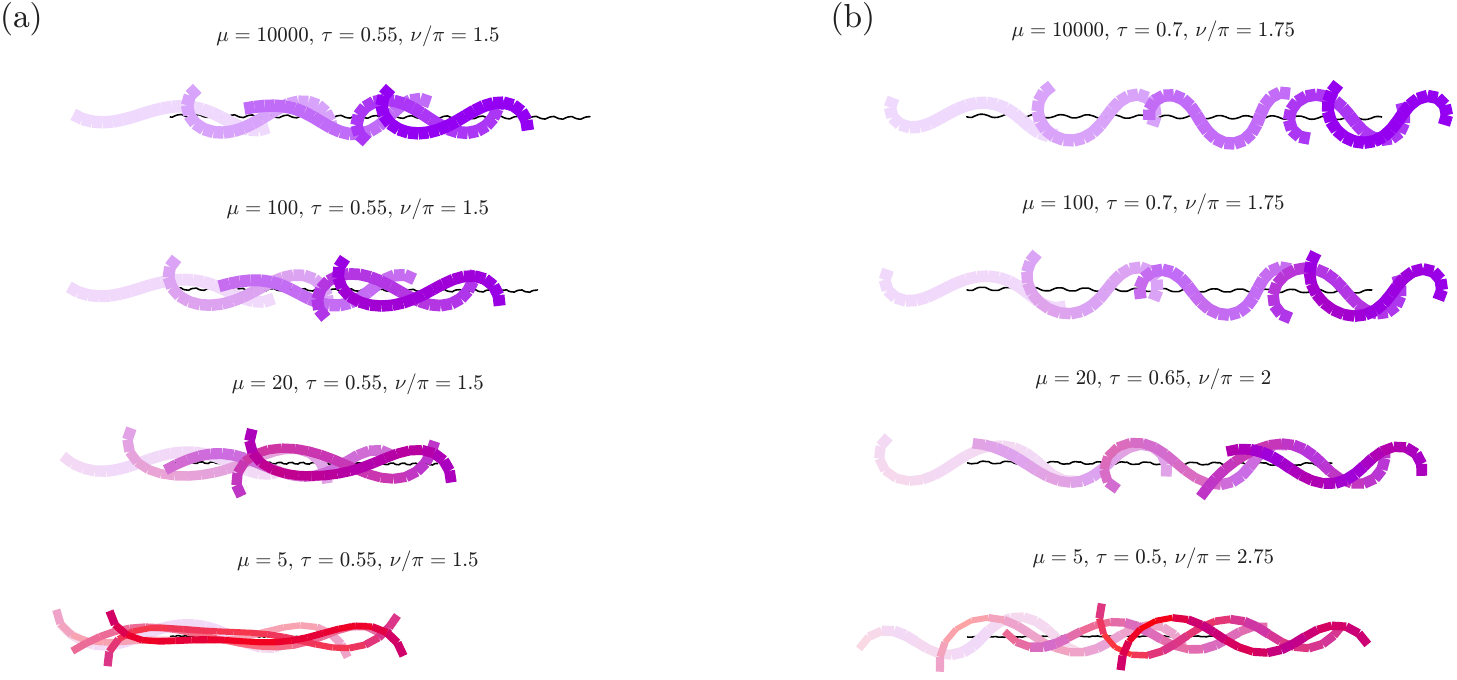}
    \caption{Effect of compressibility on bending-based locomotion. Each simulation represents the filament initially at rest on the left hand-side, actuated with a preferred curvature wave. The swimming trajectory is displayed as a black line for the displacement of the geometric center and selected snapshots for the swimmer shape. Thickness and color of the line in the filament snapshots indicate the local state of compression (purple to red, and thin to thick, from minimal to maximal compression state). The wave amplitude $\tau$ and wavenumber $\nu$ have been set arbitrarily on panel (a), showing a sharp decrease in swimming velocity for this set of parameters as passive extension stiffness is decreased from top ($\mu = 10^4$) to bottom ($\mu = 5$). In panel (b), $\tau$ and $\nu$ are the ones realising maximal forward velocity for each value of $\mu$, but with a similar velocity decrease observed as $\mu$ decreases.}
    \label{fig:passive_swimming_snapshots}
\end{figure}

\begin{figure}
    \centering
    \begin{overpic}[width=0.49\linewidth]{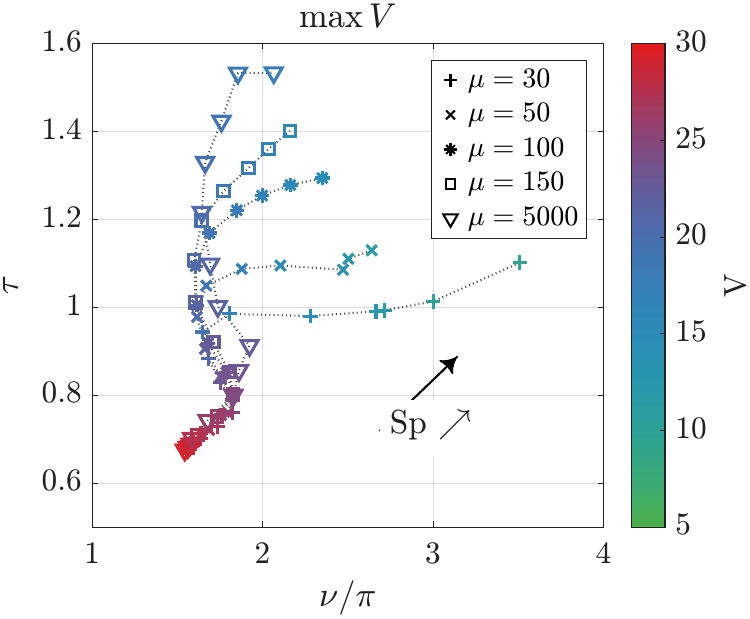}
    \put(0,80){(a)}
    \end{overpic}
    \begin{overpic}[width=0.49\linewidth]{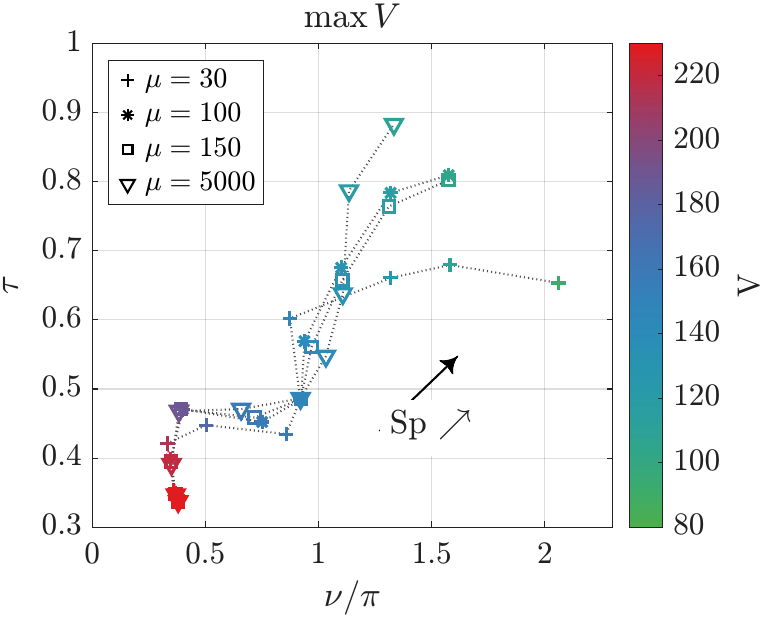}
    \put(0,80){(b)}
    \end{overpic}
    \caption{Effect of compressibility on flagellar locomotion. Both panels display points with coordinates $(\tau_{\max}(\mathrm{Sp},\mu,\gamma),\nu_{\max}(\mathrm{Sp},\mu,\gamma) )$, defined in the text as amplitude and wavenumber of curvature activity that maximise swimming velocity $V_{\max}$ for a given set of parameters $(\mathrm{Sp},\mu,\gamma)$. Each data point is coloured according to the value of $V_{\max}$ as indicated by the colorbars on the right of each plot. Each investigated value of the compression parameter $\mu$ is associated with a different marker, as indicated by the legend. For each $\mu$, ten points are shown and connected by a dotted line, corresponding to increasing integer values of $\mathrm{Sp}$ between 2 and 11, with lowest values of $\mathrm{Sp}$ at the bottom left corner and highest towards the top right corner of each panel.  (a) Swimming ($\gamma = 1/2$) (b) Crawling ($\gamma = 1/70$)}
    \label{fig:passive_swim}
\end{figure}

To study the effect of compressibility on flagellar locomotion, we first consider the classical problem of a bending-driven filament \cite{ishimoto2018elastohydrodynamical,moreau2018asymptotic} and investigate the effect of passive compressibility. To do so, we apply simple sinusoidal travelling waveforms of preferred curvature defined by \eqref{eq:alpha0_wave} -- recalling that $\omega_b$ is now equal to $1$ after nondimensionalisation. By contrast, we consider only passive compression: $\tau_e = 0$ in Eq. \eqref{eq:ell0_wave}. In this situation, compression effects will only emerge through coupling from the hydrodynamic drag forces exerted on the filament when it starts bending.

For fixed values of $\mathrm{Sp}$, $\mu$, and $\gamma$, we simulate the filament motion over 100 curvature cycles and compute the mean locomotion velocity $V$ from the displacement of the geometric center between the initial and final times. Figure \ref{fig:passive_swimming_snapshots}a shows example trajectories across values of $\mu$, with the net displacement strikingly decreasing as $\mu$ decreases in more compressible filaments. Observation of the filament behaviour on the figure suggests that this drop in swimming velocity is due to extension of the filament under forced bending, which disorganises the travelling wave structure. 

For better comparison, seeking a measure of optimal performance over the waveform parameters for a given value of $\mu$, we define $\tau_{\max}(\mathrm{Sp},\mu,\gamma)$ and $\nu_{\max}(\mathrm{Sp},\mu,\gamma)$ as the values of bending activity $\tau_b$ and bending wavenumber $\nu_b$ that achieve maximal velocity $V_{\max}(\mathrm{Sp},\mu,\gamma)$ over the range of simulated parameters: $\tau_b \in I_{\tau} = [0,2]$, $\nu_b \in I_{\nu} = [0,4\pi]$:
\begin{equation}
    V_{\max}(\mathrm{Sp},\mu,\gamma) = \max_{\tau_b \in I_{\tau}, \nu_b \in I_{\nu}}V, \qquad (\tau_{\max}(\mathrm{Sp},\mu,\gamma),\nu_{\max}(\mathrm{Sp},\mu,\gamma) ) = \argmax_{\tau_b \in I_{\tau}, \nu_b \in I_{\nu}} V.
\end{equation}
Figure \ref{fig:passive_swimming_snapshots}b shows the results for $\mathrm{Sp} = 8$, $\gamma = 1/2$ and various values of $\mu$. Even using these optimal waveform parameters, the most compressible filaments are still significantly slower than the inextensible one. 

To further confirm this observation across different $\mathrm{Sp}$ values, in the \textit{swimming case} defined as $\gamma = 1/2$, we have plotted in Figure \ref{fig:passive_swim}a the coordinates of $(\tau_{\max},\nu_{\max})$ in the $(\tau_b,\nu_b)$ plane for various values of $\mathrm{Sp}$ ranging from 2 to 11 and $\mu$ ranging in $[6,5000]$. Each point is coloured according to the value of the swimming velocity $V_{\max}$. 
The highest velocities are obtained for small values of $\mathrm{Sp}$ and for a bending wavenumber close to $1.5\pi$, consistent with classical results on optimal flagellar waveforms \cite{eloy2012kinematics}.

Figure~\ref{fig:passive_swim}a shows that the largest swimming velocities are obtained for the largest values of $\mu$, corresponding to the least compressible filaments. 
Moderate compressibility may occasionally increase $V_{\max}$ by a few percent. This variation is, however, too small to be considered significant compared to the possible impact of non-local hydrodynamic effects or nonlinear elasticity, that are neglected in our model. 
By contrast, the loss of swimming velocity observed for more extensible filaments is robust across values of $\mathrm{Sp}$ and can reach 60--70\%. Thus, at fixed $\mathrm{Sp}$, a less stretchable filament is systematically more favourable for fast bending-based swimming.

The same conclusion holds for $\gamma = 1/70$ (Figure~\ref{fig:passive_swim}b), which models crawling. Introducing passive compressibility almost always decreases the maximal locomotion velocity $V_{\max}$. Although not shown here, similar qualitative and quantitative trends are obtained for other waveforms, including partial actuation of the anterior half of the filament with the posterior half remaining passive. Overall, passive extensibility slows the swimmer in the overwhelming majority of parameter sets.

Thus, passive compressive elasticity alone does not provide a mechanism for improving flagellar swimming in this setting. Across the parameter ranges considered here, extensibility either has a negligible effect or reduces the maximal velocity, with the largest losses observed for the most compressible filaments. This suggests that any beneficial role of compression in flagellar locomotion must arise from active compressive actuation or from more specific mechanical couplings, rather than from passive extensibility alone.

\subsection{Active compression}

\begin{figure}
    \centering
    \includegraphics[width=\linewidth]{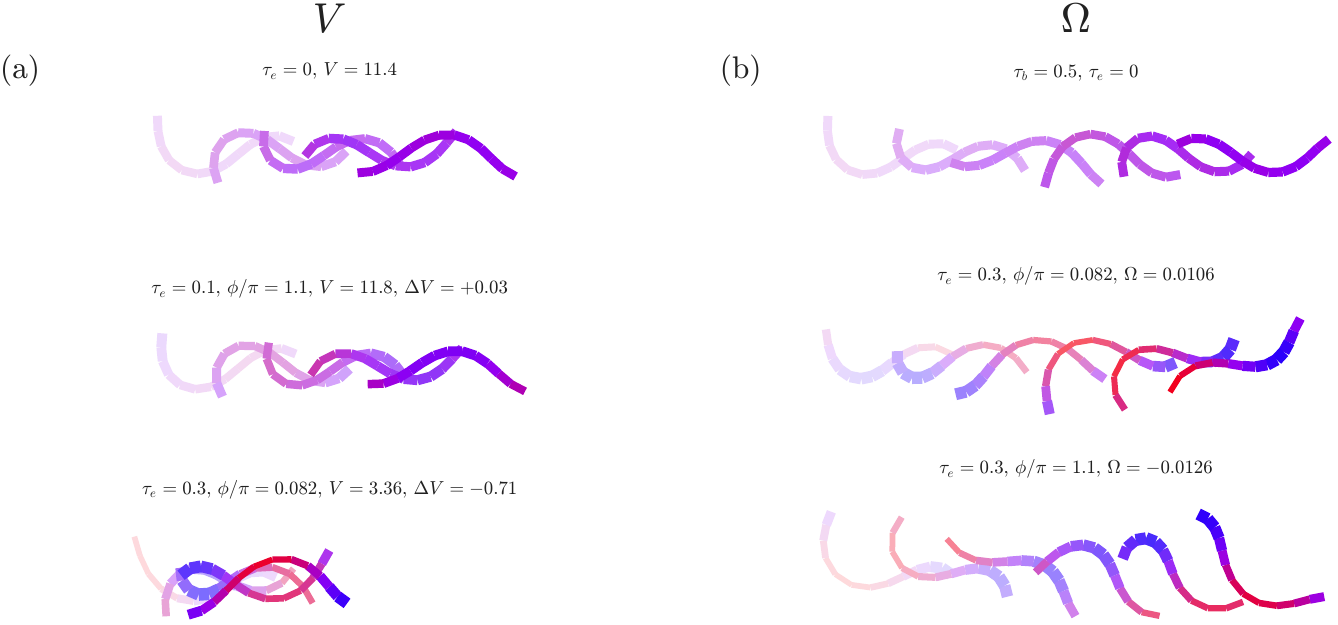}
    \caption{Effect of active compression on bending-based locomotion. Each simulation represents the filament actuated with a fixed preferred curvature wave ($\tau_b = 0.5$), combined with a preferred elongation wave varying between each panel. (a) Frequency of active elongation is twice the frequency of active bending, ensuring no net rotation. From top to bottom: no active compression; highest increase in velocity for active compression across $\tau_e \in [0,0.3]$, $\nu_e = 1$ and $\phi \in [0,2\pi]$; highest decrease across the same parameter range. (b) Frequency of bending and compression are equal, generating net rotation. For better readability, the filament snapshots are separated horizontally, while almost no horizontal motion occurs in the fixed frame. From top to bottom: no active compression, highest left and right rotation across the tested parameter range.}
    \label{fig:act_comp_snaps}
\end{figure}

\begin{figure}
    \centering
    \includegraphics[width=0.83\linewidth]{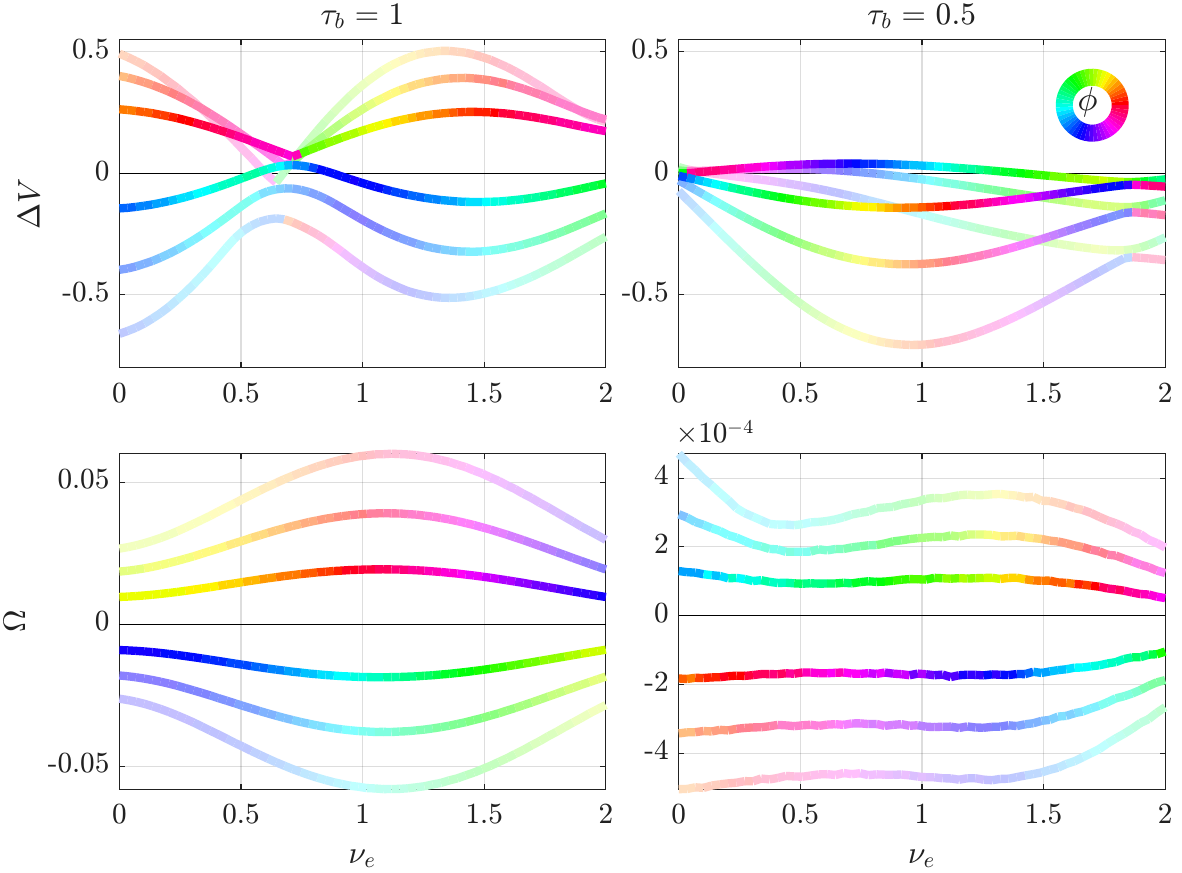}
    \caption{{Effect of active compression on bending-driven swimming (top row) and rotation (bottom row). For a travelling wave of active compression with wavenumber $\nu_e$, varying on the $x$-axis of each panel, the values on $y$-axis indicate the minimal and maximal values of $\Delta V$ (top row) and $\Omega$ (bottom row) obtained when varying the phase shift $\phi$. The amplitude of the compression wave $\tau_e$ takes values 0.1 (vivid color line), 0.2 (lightly faded color), 0.3 (strongly faded color). The color of each data point indicates the value of $\phi_{\max}$, the phase shift at which the corresponding value of $\Delta V$ or $\Omega$ is reached. For example, on the top left panel, for $\nu_e = 1$, looking at the vivid color line: $\Delta V_{\min} \approx -0.05$ and $\Delta V_{\max} \approx 0.15$, meaning that for $\tau_e = 0.1$, active compression wave can decrease swimming velocity $V$ by up to 5\% and increase it by up to 15\%. Moreover, still at $\nu_e=1$, the line for $\Delta V_{\min}$ is blue and the line for $\Delta V_{\max}$ is light green, meaning that these minimal and maximal velocity changes occur at approximately $\phi = -\pi/2$ and $\phi = \pi/2$, respectively.}}
    \label{fig:active}
\end{figure}

Having established that passive compressibility does not generally improve swimming, we now consider active bending-compression coupling, with compressive actuation prescribed through Eq.~\eqref{eq:ell0_wave}.

To assess the contribution of active compression, we fix parameters that give efficient locomotion in the nearly inextensible case, based on the passive-compression sweep above. We focus on swimming ($\gamma=1/2$) with $\mathrm{Sp}=5$ and $\nu_b=1.6\pi$. We consider two bending amplitudes: $\tau_b=1$, close to the value producing optimal swimming in the passive case, and $\tau_b=0.5$, which yields smaller deformations qualitatively closer to biological flagella such as spermatozoa. Finally, we fix $\mu=20$, corresponding to a regime of large compressibility.

Previous kinematic theory \cite{ishimoto2025bending} showed that a compression frequency twice the bending frequency preserves swimming in a straight line, whereas equal bending and compression frequencies can generate curved trajectories. We therefore consider a forward-swimming case with $\omega_e=2$ and a turning case with $\omega_e=1$. For each case, we use compression amplitudes $\tau_e=0.1$, $0.2$, and $0.3$, wavenumbers from uniform compression, $\nu_e=0$, to twice the bending wavenumber, and phase shifts $\phi$ from $0$ to $2\pi$.

For each waveform, we compute the swimming velocity $V$ and the relative change
\[
    \Delta V(\tau_e,\nu_e,\phi) = \frac{V-V_0}{V_0},
\]
where $V_0$ is the swimming velocity without active compression. In the case where the filament is turning, we use the rotational velocity $\Omega$ as compared to the passive compression case, where $\Omega=0$.

Figure \ref{fig:act_comp_snaps} illustrates the effect of active compression on two examples of straight swimming and turning. Panel (a) shows that active compression can increase swimming velocity by a few percent or decrease it drastically, depending on the phase shift $\phi$. Similarly, as seen on panel (b), left or right turning can occur depending on the phase shift between bending and compression waves. 

To quantitatively assess the effect of active compression, for fixed $\tau_e$ and $\nu_e$, we define $(\phi_{\max},\Delta V_{\max})$ and $(\phi_{\min},\Delta V_{\min})$ as the phase shifts and corresponding extrema of $\Delta V$ over $\phi \in [0,2\pi]$. 

The numerical results are shown in Figure~\ref{fig:active}. Depending on the compression wavenumber $\nu_e$, active compression can have substantial effect on swimming velocity. In the forward-swimming case with larger bending amplitude (top left panel), the strongest positive and negative changes occur for uniform compression, $\nu_e=0$, and around $\nu_e/\nu_b \approx 1.3$. For these wavenumbers, a suitable phase shift changes $V$ by approximately $-10\%$ to $20\%$ for weak compression, $\tau_e=0.1$, and by as much as about $50\%$ for the strongest compression considered here, $\tau_e=0.3$. Conversely, choosing $\nu_e/\nu_b \approx 0.6$ greatly reduces the effect of active compression, which then modifies $V$ by only a few percent. For the smaller bending amplitude (top right panel), active compression mostly decreases the swimming velocity. 

The bottom row shows that active compression can generate rotation when coupled to an otherwise forward-swimming bending gait, as expected from \cite{ishimoto2025bending}. The largest rotational velocities occur when the compression and bending wavenumbers are comparable, $\nu_e/\nu_b \approx 1$. Importantly, net rotation is generated here through coupling between bending and compression even though the prescribed deformation modes are individually symmetric over one cycle. Under this symmetry constraint, bending alone or compression alone would not generate a net rotational bias.

\section{Follower-force model}
\label{sec:follower}

\begin{figure}
    \centering
    \includegraphics[width=0.66\linewidth]{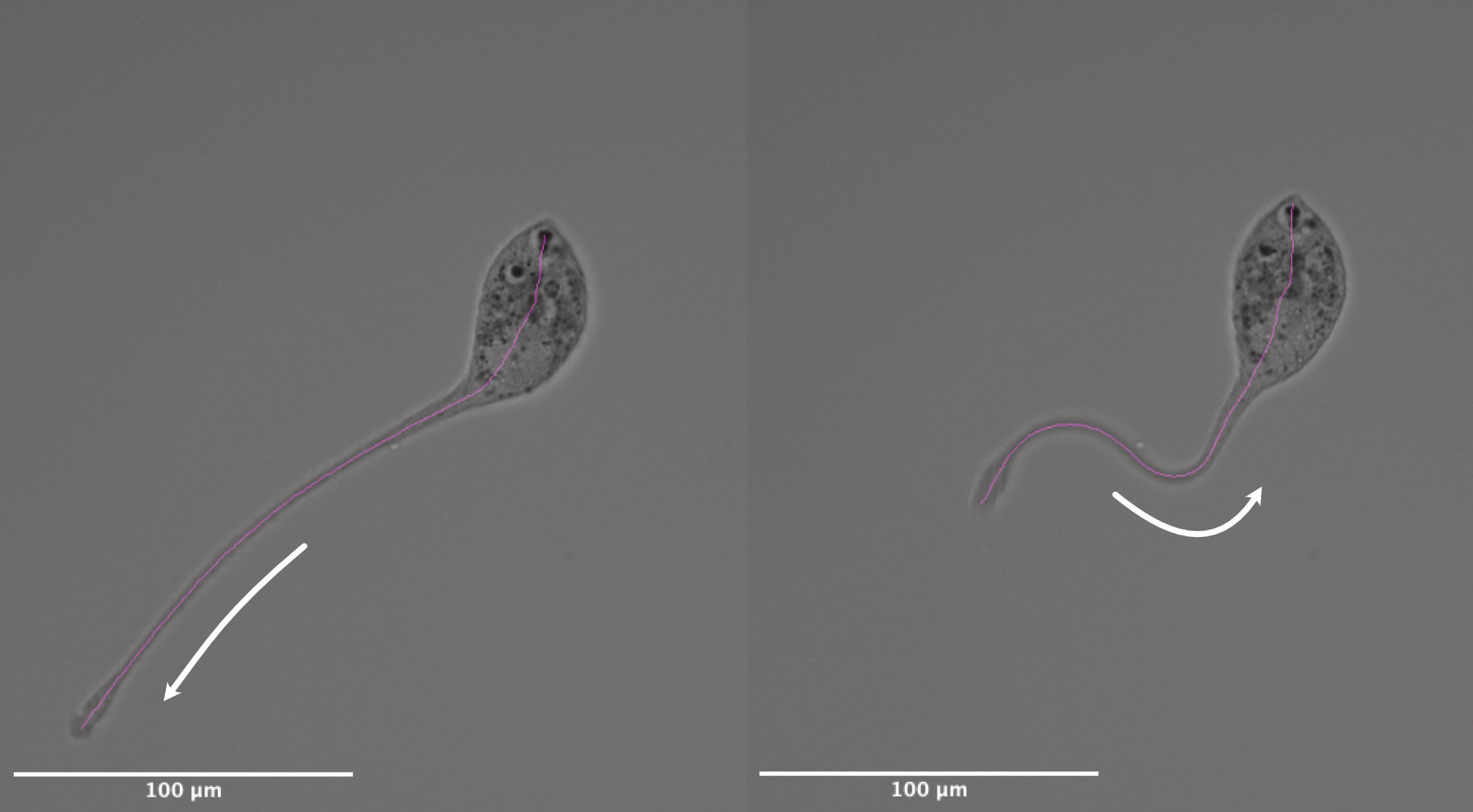}\\ \medskip
    \includegraphics[width=0.85\linewidth]{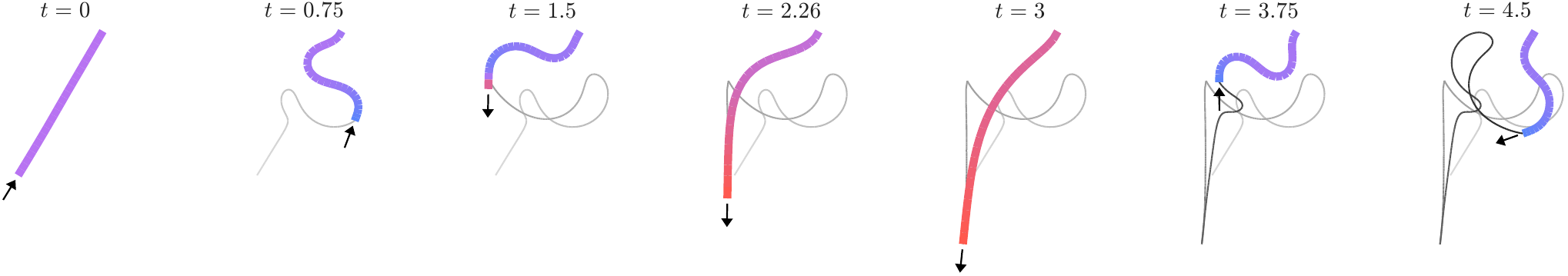}
    \caption{Representative \textit{Lacrymaria olor}-like length-changing dynamics. The experimental image sequence illustrates large neck contraction with a time interval of $0.2\,\mathrm{s}$ between the two images. The skeletonized shape is overlaid for illustration.
    The lower panel shows a follower-force simulation with $N=24$, $\mathrm{Sp}=7$, $\mu=10^2$, and $A=0.95$.}
    \label{fig:lacry}
\end{figure}

\subsection{Lacrymaria: a model organism for compression-based behaviour}

In this section, we use \textit{Lacrymaria olor} and other highly extensible microorganisms as motivation to assess how large compressibility can affect exploration and locomotion behaviours. The behaviour of \textit{Lacrymaria olor} has recently attracted renewed attention. Its large body and neck deformations involve several competing time scales \cite{coyle2019coupled,qin2024dynamic}, making it a natural test case for compressible elastohydrodynamics.

\textit{Lacrymaria} alternates between extension and contraction of its neck. During contraction, the neck can undulate, and it straightens during extension, as illustrated in Figure~\ref{fig:lacry} (See Appendix~\ref{app:expr} for the details of cell culture and observation). Krishnamurthy and Prakash \cite{krishnamurthy2023emergent} proposed a minimal model, referred to as the KP model in the remainder of this study, for this behaviour based on a clamped filament driven by a time-periodic tangential follower force. In the pushing phase, the filament buckles and oscillates, while the distal end is drawn back and reoriented in the pulling phase. Depending on the activity amplitude and forcing period, this model transitions from simple periodic motion to higher-period and apparently chaotic behaviours, suggesting a mechanical route to exploratory dynamics.

The KP model is based on an $N$-bead representation with linear and angular springs, which is closely related in spirit to our $N$-segment formulation. However, the simulations reported in that work involve only modest length variation. In addition, the distributed hydrodynamic drag along each segment in the present formulation may become important when the filament undergoes large length changes. We therefore use the compressible $N$-link model to ask how axial compliance modifies the emergent behaviour of a filament driven by the same type of push-pull follower force.

In the following sections, we first compare the clamped model with the behaviours reported in the KP model in \cite{krishnamurthy2023emergent}, then vary $\mu$ to assess how extensibility affects periodic and aperiodic dynamics. In particular, we study how compressibility modifies the behaviour qualified as \textit{escape} in the KP model, where the filament irreversibly bends to one side at high activity.  
Motivated by \textit{Lacrymaria}-like locomotion, we then remove the clamp and perform a similar parameter study for a free filament driven by the same periodic follower force.

\subsection{Clamped filament for exploration}

\begin{figure}
    \centering
    \includegraphics[width=\linewidth]{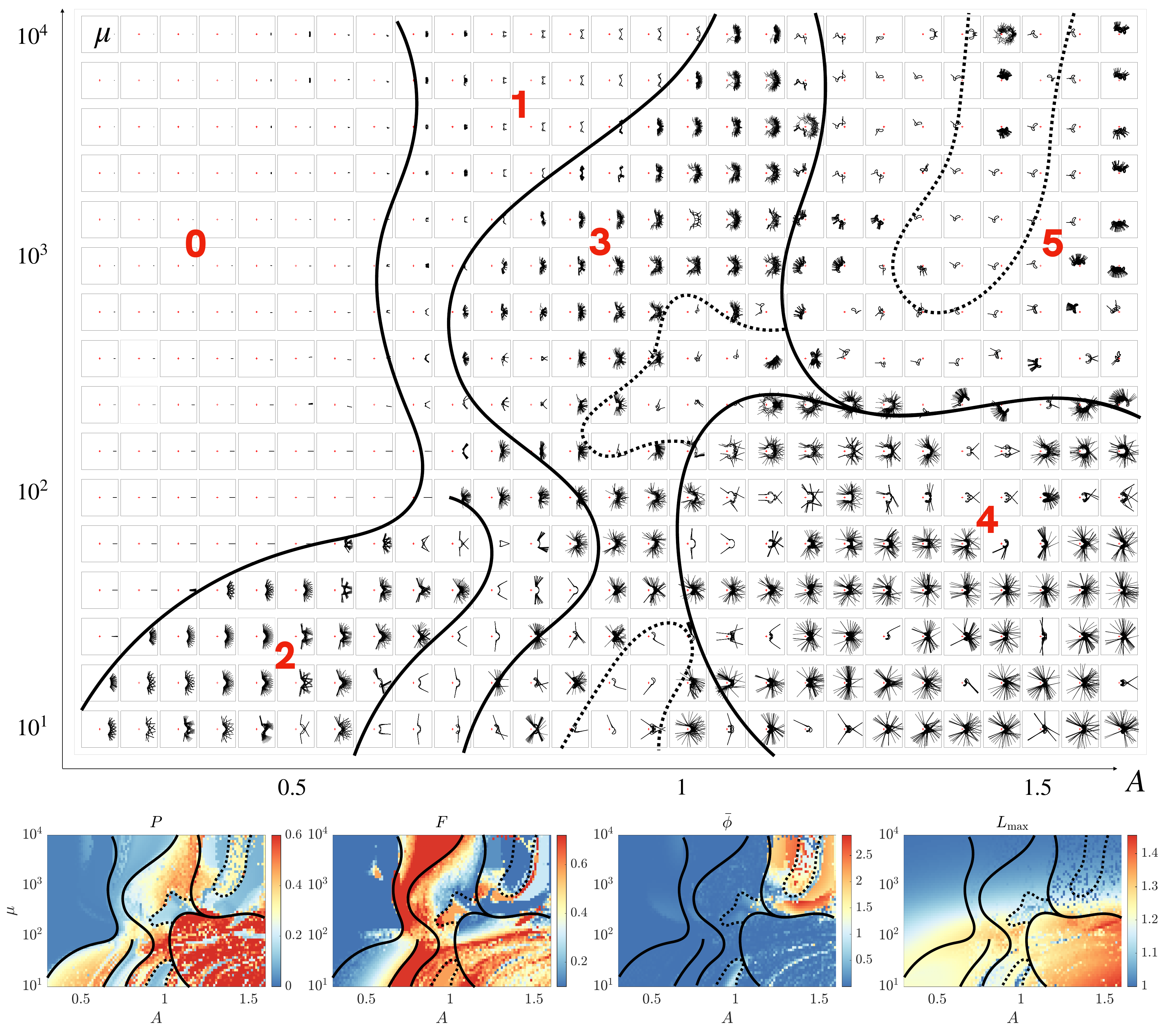}
    \caption{Clamped follower-force dynamics as a function of activity amplitude $A$ and extensibility parameter $\mu$. The upper phase diagram shows the trajectory of the filament head after transients have been removed; the clamped endpoint is marked in red. The lower panels report the quantitative metrics $P$, $F$, $\bar{\phi}$ and $L_{\max}$ used to identify transitions between periodic, alternating, aperiodic, and escape-like behaviours.}
    \label{fig:clamp}
\end{figure}

Mimicking the KP model, we apply a tangential force to the distal tip of the $N$-segment filament according to Eq.~\eqref{eq: follower force}. The force amplitude is prescribed as the square wave
\begin{equation}
    F_{\mathrm{act}} (t) = A \;\mathrm{sign} (\cos(\omega_{\mathrm{ff}} t)). 
\label{eq: follower periodic}
\end{equation}

We fix $\omega_{\mathrm{ff}}=1.4$, which approximately corresponds to $\tau=1700$ in the study by Krishnamurthy and Prakash and produces a range of behaviours, including transitions from periodic to aperiodic motion, as the activity amplitude $A$ varies from 0 to $1.5$. We also vary $\mu$, which characterises the axial extensibility of the filament. For the active forces considered here, values $\mu \gtrsim 10^4$ correspond to an effectively inextensible filament, whereas $\mu \approx 10$ allows very large extension of the distal segments where the follower force is applied, up to approximately 700\%. For very small $\mu$, we add a short-range elastic resistance when any segment length falls below 10\% of its rest length, preventing non-physical collapse to zero length. Finally, we take $\mathrm{Sp}=5$, matching the estimated value of $\rm Sp$ from the KP model. 

Figure~\ref{fig:clamp} summarises the numerical study. To illustrate the richness of the emergent behaviour, each small panel shows the trajectory of the head after removing the first 10 cycles to eliminate transients. The head trajectory is shown as a black line, and the clamped endpoint is marked by a red dot. For $\mu=10^4$, the filament is nearly inextensible and the simulations reproduce the qualitative sequence reported for the KP model. A transition to buckling occurs near $A=0.5$, followed by two-periodic cycles and then by more aperiodic motion around $A=1$. Beyond $A=1.2$, an ``escape'' state dominates, in which the filament remains preferentially bent to one side rather than alternating between sides.

Both the filament behaviour and the transition values change strongly with $\mu$, as seen in Figure~\ref{fig:clamp}. To organise these behaviours, we define four quantitative metrics, plotted below the phase diagram.

The explored fraction $P$ measures how much of a unit circle around the clamped endpoint is visited by the filament head over the simulation. A small value, of order $0.1$, indicates that the head repeatedly visits the same locations, typically in a one- or two-periodic state. Intermediate values, around $0.3$--$0.4$, indicate that the filament explores a larger range of orientations, as occurs when aperiodic behaviour emerges, even if lateral excursions remain moderate. The largest values of $P$ correspond to trajectories that explore both the forward region and the lateral sides.

The alternating fraction $F$ measures how regularly the filament changes side after each actuation cycle. Low values indicate that the head remains on the same side for many cycles; high values, here roughly above $0.6$, indicate regular alternation; intermediate values indicate erratic switching. The third metric, $\bar{\phi}$, is the average head orientation at the end of each actuation cycle. It is useful for identifying escape states, in which the filament develops a persistent bias to one side. Finally, $L_{\max}$ measures the maximum filament length over the simulation and directly reports the magnitude of extension generated by high activity and low axial stiffness. Together, these metrics identify transitions that are represented by lines in Figure~\ref{fig:clamp} and reported on the main phase diagram. This yields the qualitative regions summarised in Table~\ref{tab:clamp_categories}.

\begin{table}
    \centering
    \begin{tabular}{c|c|c|c|c}
         Reg. & $P$ & $F$ & $\bar{\phi}$ & $ L_{\max}$ \\ \hline
         0    &  low & low & low & low  \\
         1    &  low & high & low & var \\
         2    &  med & low & low & med \\
         3    &  med & med & low & var \\
         4    &  high & med & low & high \\
         5    & low & low & high & low     
    \end{tabular}
    \caption{Characterisation of the six regions in the phase space of Figure \ref{fig:clamp}.}
    \label{tab:clamp_categories}
\end{table}

Region 0 in Figure~\ref{fig:clamp} corresponds to parameters for which little motion is visible because the follower force is too weak to trigger buckling. In the nearly inextensible case, buckling and subsequent oscillation start around $A=0.6$. This critical value remains relatively stable until $\mu$ approaches $10^2$, where a distinct behaviour emerges (Region 2). Because the filament is then more extensible, the pulling phase does not fully realign the filament. Curvature can accumulate away from the tip during successive pushing phases, producing slowly beating motion even at weak activity.

Increasing activity brings the filament into Region 1, which corresponds to simple periodic trajectories in which the filament alternates symmetrically between the left and right sides. The beating period is therefore twice the forcing period. Above a threshold that depends non-monotonically on $\mu$, the filament displays more complex dynamics, ranging from longer-period motion to apparently aperiodic behaviour (Region 3). This matches the observations for the KP model in the inextensible case.
For nearly inextensible filaments, activity larger than approximately $A=1.1$ produces a final transition to Region 5, where the filament usually experiences an escape state, persistently beating to one side. The escape state is characterised by beating with the same period as the forcing, returning to approximately the same point after each actuation cycle. In the KP model, this was interpreted as unsuitable for exploration, because the filament becomes confined to one side.

In the high-activity regime, increasing extensibility can suppress escape and recover aperiodic behaviour (Region 4), similar to Region 3. These trajectories in Region 4 differ, however, from those in Region 3 because the filament reaches farther to the sides during each activation cycle, producing larger values of the explored fraction $P$. In this sense, axial compliance can help the filament explore regions around its base without becoming trapped on one side.

\subsection{Free filament for locomotion}

\begin{figure}
    \centering
    \includegraphics[width=0.66\linewidth]{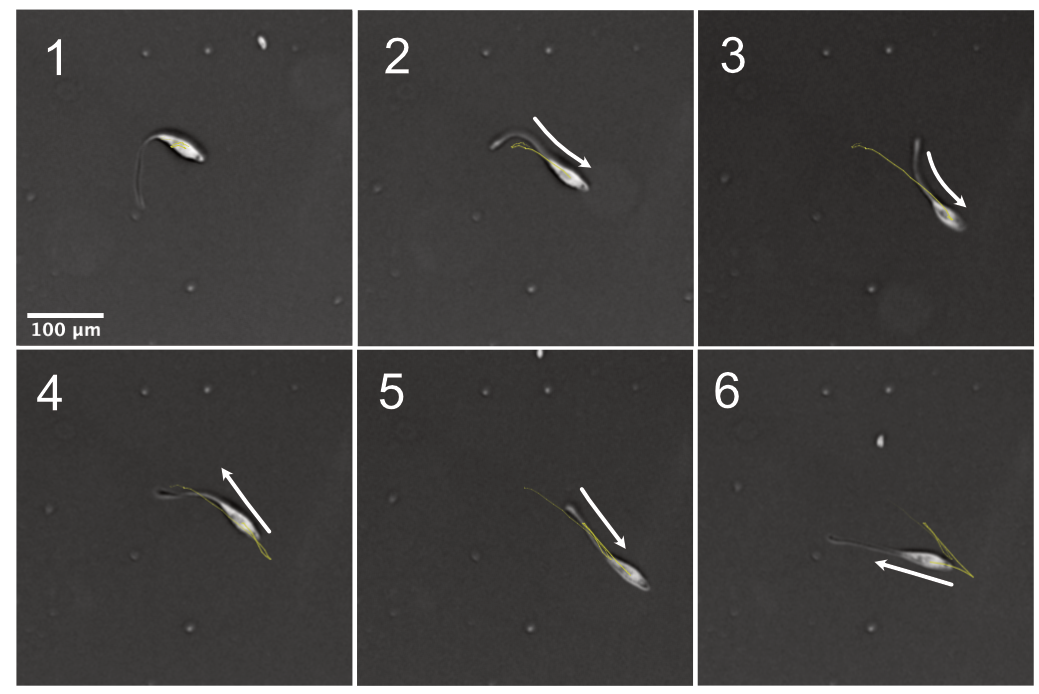}
    \includegraphics[width=0.62\linewidth]{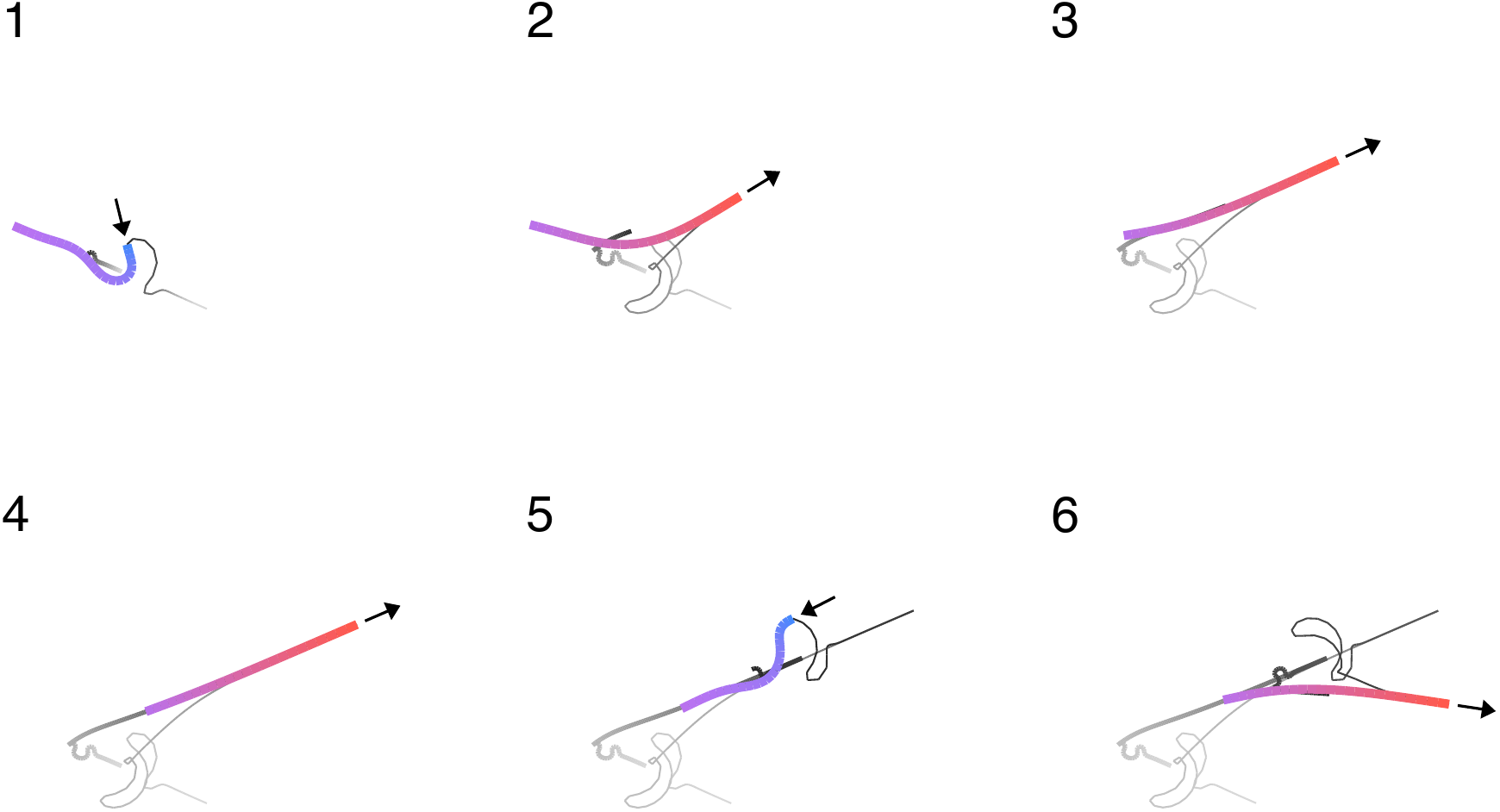}
    \caption{Microscopy image of \textit{Lacrymaria} during swimming and simulation of free-end follower force filament. In the top panel, images are shown at intervals of 0.39\,s from frame 1 to frame 6. Yellow lines show trajectories of the body centre during the preceding 1.72\,s. The white arrow schematically indicates the instantaneous velocity of the body. In the bottom panel, the dynamics of a free filament with periodic follower force is shown at selected snapshots, with the buckling-induced reorientation mechanism seen on Panels 5-6. On each panel, the thin and thick gray lines show the trajectory of the tip and the geometric center of the filament, respectively.}
    \label{fig:lacry-swim}
\end{figure}

\begin{figure}
    \centering
    \includegraphics[width=\linewidth]{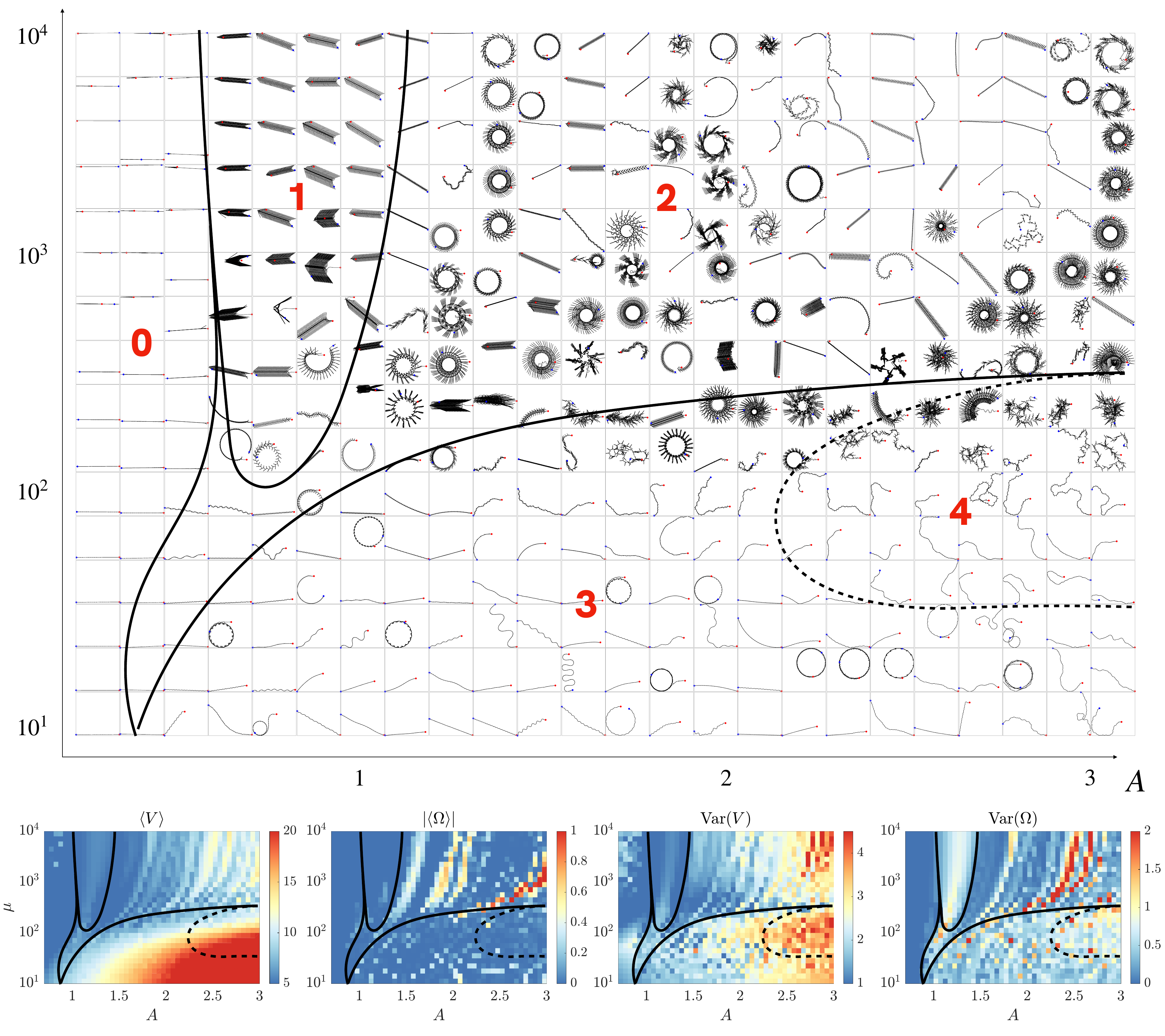}
    \caption{Free-filament follower-force dynamics as a function of activity amplitude $A$ and extensibility parameter $\mu$. The upper phase diagram shows the trajectory of the filament geometric center; initial and final points are indicated by a red and blue dot, respectively. The lower panels report the quantitative metrics used to identify transitions between behaviours: mean and variance of the translational velocity $V$ and rotational velocity $\Omega$.}
    \label{fig:swim}
\end{figure}

The previous section considered a clamped filament, which can generate exploration-like motion when driven by an oscillatory follower force. 
Motivated by \textit{Lacrymaria} and other organisms that exhibit oscillatory back-and-forth motion coupled to body compression, as illustrated in Figure~\ref{fig:lacry-swim}, we now remove the clamp and ask whether the same follower-force forcing used in the previous section can generate locomotion and reorientation of a free filament. 

We use the same periodic follower-force law \eqref{eq: follower periodic}, but impose free boundary conditions at both ends. In this configuration, the filament is pulled through space during one phase of the forcing and buckles while moving during the other. Depending on the duration and amplitude of the buckling phase, this can generate periodic or aperiodic changes of direction, analogous to the clamped dynamics above but now coupled to rigid-body translation and rotation. 

We use the same parameter values as in the clamped study, except that the activity amplitude $A$ is increased up to 3 to access a wider range of behaviours. The results are shown in Figure~\ref{fig:swim}. Each small panel shows the trajectory of the filament geometric centre over 60 actuation cycles. The starting and ending points are marked by red and blue dots, respectively.

Visual inspection of the phase diagram reveals several qualitative classes of trajectories: regular straight trajectories corresponding to forward locomotion, regular circular trajectories, and irregular trajectories ranging from slightly curved paths to apparently diffusive exploration.

To characterise these behaviours, we compute the translational velocity $V$ and rotational velocity $\Omega$ after each cycle, giving two time series for each trajectory. We then compute the means $\langle V\rangle$, $\langle\Omega\rangle$ and variances $\mathrm{Var}(V)$, $\mathrm{Var}(\Omega)$. These four metrics, shown below the phase diagram in Figure~\ref{fig:swim}, provide a coarse classification into qualitative behavioural regions, summarised in Table \ref{tab:swim_categories}.

\begin{table}
    \centering
    \begin{tabular}{c|c|c|c|c}
         Reg. & $\langle V \rangle$ & $|\langle \Omega \rangle|$ & $\mathrm{Var}(V)$ & $\mathrm{Var}(\Omega)$ \\ \hline
         0    &  low & low & low & low  \\
         1    &  low & low & low & med \\
         2    &  med & high* & var & high* \\
         3    &  high & low & med & med \\
         4    &  high & low & high & med \\   
    \end{tabular}
    \caption{Qualitative characterisation of the regions in the free-filament phase diagram of Figure~\ref{fig:swim}. The asterisks in Region 2 row indicate that high values are not observed uniformly within the region.}
    \label{tab:swim_categories}
\end{table}

Region 0, at low activity, shows no buckling and therefore little net behaviour beyond back-and-forth motion. Region 1, at intermediate activity and low extensibility, corresponds to periodic trajectories with alternating motion to the left and right, producing forward locomotion on average. Region 2 appears at high activity and low extensibility. The behaviours in this region are diverse and difficult to describe with a single metric, but many trajectories are periodic and biased to one side, producing circular locomotion on average. Depending on activity and extensibility, however, forward locomotion or aperiodic motion may also occur within this region.

Aperiodic behaviour is more prevalent in Regions 3 and 4, where the high variances of $V$ and $\Omega$ indicate the loss of simple periodicity. In particular, a larger variance of $V$ corresponds to trajectories with larger changes in local curvature, and therefore to more exploratory motion. Region 4 marks the parameter range where $\mathrm{Var}(V)$ is largest, suggesting an enhanced exploration capacity induced by moderate compressibility.

\section{Discussion}
\label{sec:discussion}

We have introduced a compressible $N$-link elastohydrodynamic model in which segment lengths and bending angles evolve as coupled dynamical variables. This formulation preserves the finite-dimensional structure of classical $N$-link swimmers while allowing large axial deformation. It therefore provides a practical route to studying active slender bodies for which length change plays a significant role.

The bending-driven locomotion results in Section \ref{sec:locomotion} show that passive extensibility alone is not a robust mechanism for enhancing swimming or crawling. In most of the parameter space explored here, axial compliance reduces locomotion speed, presumably because part of the actuation is absorbed into length change rather than converted into shape changes that generate propulsion. This negative result is useful because it separates passive compliance from active length control: if compression improves performance, it is likely to do so through active phase-dependent coupling rather than through passive softening alone.

Active compression produces a richer picture. Depending on the wavelength, frequency, and phase of the rest-length wave, compression can enhance forward swimming, reduce it, or generate turning. This agrees qualitatively with kinematic theories of coupled bending and compression \cite{ishimoto2025bending}, while showing that similar effects persist when the filament shape is determined dynamically by elasticity and hydrodynamic drag.

The follower-force examples in Section \ref{sec:follower}, motivated by the observed behaviour of \textit{Lacrymaria}, highlight another role for compressibility. In clamped filaments, axial compliance changes the onset and organisation of buckling, aperiodicity, and escape-like states. In free filaments, the same push-pull actuation can generate translation, rotation, and irregular exploration-like trajectories. These results suggest that length change can tune the conversion of a simple periodic input into complex spatial behaviour, a mechanism that may be relevant to organisms such as \textit{Lacrymaria}.

Several model improvements can be considered to make the idealised model of a slender filament studied here more realistic. The resistive-force-theory model for hydrodynamic drag neglects long-range hydrodynamic interactions, which may be important when strongly compressed configurations bring distant filament portions close together. The follower-force model inspired by \textit{Lacrymaria} and used in Section~\ref{sec:follower} is known to represent ciliary activity relatively well, but it is still a reduction to a tip force only, whereas real organisms involve distributed ciliary forcing and complex internal bending actuation through molecular motors. Finally, motivated by longer-time length regulation in living systems such as \textit{Lacrymaria} \cite{qin2024dynamic,krishnamurthy2023emergent}, one could consider letting the rest lengths evolve on a slower time scale, for example through a viscoelastic relaxation law $\dot{\ell}_{0,i} = k_v(\ell_i-\ell_{0,i})$.

Overall, our results highlight the distinct roles of passive compliance and active length change in slender-body locomotion. While passive extensibility generally hinders bending-driven propulsion, active compression provides an additional mechanism for controlling swimming speed and direction. Under follower-force actuation, axial compliance can also profoundly alter the nonlinear dynamics of the filament, promoting transitions between periodic, irregular, and exploratory behaviours. Beyond their relevance to biological locomotion, these findings point towards new opportunities in the design of soft microrobots, where controlled compression and bending could be exploited to achieve versatile locomotion and adaptive behaviours.

\section*{Acknowledgements}
We thank Dr. Atsushi Taniguchi (Hokkaido University) for collecting the \textit{Lacrymaria olor} cells used in this study and maintaining the cultures. K.~I. acknowledges JSPS KAKENHI (Grant No. 24K21517) and the Japan Science and Technology Agency (JST), FOREST (Grant No. JPMJFR212N) and CREST (Grant No. JPMJCR25Q1). Y. N. acknowledges JSPS KAKENHI (Grant No. 26H02509).
C. M. and J. H. acknowledge funding overseen by the French National Research Agency (ANR) and France 2030 as part of the Organic Robotics Programme (PEPR O2R).
\appendix

\section{Assembly of the hydrodynamic matrix}
\label{app:matrix}

This appendix gives the explicit matrix assembly used in Eq.~\eqref{eq:linear_system}. We first write the balance equations in segment variables, then convert them to the relative coordinates used in the main text.

Let
\[
    \mathbf{q}
    =
    (x_1,\dots,x_N,y_1,\dots,y_N,\theta_1,\dots,\theta_N,
    \ell_1,\dots,\ell_N)^T
\]
collect the proximal endpoint position, orientation, and length of each segment. The hydrodynamic part of the balance equations can be written as
\[
    \mathsf{A}\dot{\mathbf{q}},
    \qquad
    \mathsf{A}
    =
    \mathrm{Sp}_{\Delta}^4
    \begin{pmatrix}
        \mathsf{A}^{F} \\
        \mathsf{A}^{T} \\
        \mathsf{A}^{E}
    \end{pmatrix},
\]
where $\mathsf{A}^{F}$ contains the two global force-balance rows, $\mathsf{A}^{T}$ contains the $N$ torque-balance rows, and $\mathsf{A}^{E}$ contains the $N$ extensional force-balance rows. The first row of $\mathsf{A}^{T}$ is the torque balance on the whole filament about $\mathbf{x}_1$; rows $2,\dots,N$ are the partial torque balances on $\{S_i,\dots,S_N\}$ about $\mathbf{x}_i$.

For compactness, set
\[
    c_i=\cos\theta_i,\qquad s_i=\sin\theta_i,\qquad
    a_{ij}=x_j-x_i,\qquad b_{ij}=y_j-y_i.
\]
The global force-balance coefficients associated with segment $j$ are
\begin{align}
    A^{F}_{x,x_j} &=
        -\ell_j(\gamma c_j^2+s_j^2), &
    A^{F}_{x,y_j} &=
        -\ell_j(\gamma-1)c_js_j, \nonumber\\
    A^{F}_{x,\theta_j} &=
        \frac{\ell_j^2}{2}s_j, &
    A^{F}_{x,\ell_j} &=
        -\frac{\gamma\ell_j}{2}c_j, \label{eq:app_force_x}\\
    A^{F}_{y,x_j} &=
        -\ell_j(\gamma-1)c_js_j, &
    A^{F}_{y,y_j} &=
        -\ell_j(c_j^2+\gamma s_j^2), \nonumber\\
    A^{F}_{y,\theta_j} &=
        -\frac{\ell_j^2}{2}c_j, &
    A^{F}_{y,\ell_j} &=
        -\frac{\gamma\ell_j}{2}s_j. \label{eq:app_force_y}
\end{align}
All other entries in the two force-balance rows are obtained by placing these coefficients in the columns corresponding to $(x_j,y_j,\theta_j,\ell_j)$ and summing over $j=1,\dots,N$ through matrix multiplication.

The torque-balance row indexed by $i$ receives contributions only from segments $j\geq i$. For $j<i$, all entries are zero. For $j\geq i$, the nonzero coefficients are
\begin{align}
    A^{T}_{i,x_j}
        &=
        \frac{\ell_j^2}{2}s_j
        +\ell_j a_{ij}(1-\gamma)s_jc_j
        +\ell_j b_{ij}(\gamma c_j^2+s_j^2), \nonumber\\
    A^{T}_{i,y_j}
        &=
        -\frac{\ell_j^2}{2}c_j
        +\ell_j b_{ij}(\gamma-1)s_jc_j
        -\ell_j a_{ij}(c_j^2+\gamma s_j^2), \nonumber\\
    A^{T}_{i,\theta_j}
        &=
        -\frac{\ell_j^3}{3}
        -\frac{\ell_j^2}{2}(a_{ij}c_j+b_{ij}s_j), \nonumber\\
    A^{T}_{i,\ell_j}
        &=
        \frac{\gamma\ell_j}{2}(-a_{ij}s_j+b_{ij}c_j).
    \label{eq:app_torque}
\end{align}
These terms are obtained by inserting the resistive-force expression into the torque integral about $\mathbf{x}_i$.

The extensional rows are force balances on the distal subsystem
$\{S_i/2,S_{i+1},\dots,S_N\}$ projected onto $\mathbf{e}_{i,\parallel}$. It is useful to express them using the force coefficients in Eqs.~\eqref{eq:app_force_x}--\eqref{eq:app_force_y}. For any variable $\xi_j\in\{x_j,y_j,\theta_j,\ell_j\}$, define
\[
    G_{i,j}^{\xi}
    =
    A^{F}_{x,\xi_j}c_i + A^{F}_{y,\xi_j}s_i.
\]
Then the extensional row $i$ has no contribution from segments $j<i$. For $j>i$,
\[
    A^{E}_{i,\xi_j}=G_{i,j}^{\xi},
\]
whereas for the half-segment contribution $j=i$,
\begin{align}
    A^{E}_{i,x_i} &= \frac{1}{2}G_{i,i}^{x}, &
    A^{E}_{i,y_i} &= \frac{1}{2}G_{i,i}^{y}, \nonumber\\
    A^{E}_{i,\theta_i} &= \frac{3}{4}G_{i,i}^{\theta}, &
    A^{E}_{i,\ell_i} &= \frac{3}{4}G_{i,i}^{\ell}.
    \label{eq:app_extension}
\end{align}
The factors $1/2$ and $3/4$ come from integrating the translational and rotational parts of the resistive-force density over the distal half of segment $S_i$. 

The matrix $\mathsf{A}$ acts on segment-coordinate velocities $\dot{\mathbf{q}}$, whereas Eq.~\eqref{eq:linear_system} is written in the relative coordinates
\[
    \mathbf{X}_{\mathrm{rel}}
    =
    (x,y,\theta,\alpha_1,\dots,\alpha_{N-1},
    \ell_1,\dots,\ell_N)^T.
\]
The conversion $\dot{\mathbf{q}}=\mathsf{C}\dot{\mathbf{X}}_{\mathrm{rel}}$ is determined by the kinematics. It is convenient to introduce
\[
    \beta_1=\theta,\qquad
    \beta_k=\alpha_{k-1}\quad (k=2,\dots,N),
\]
so that
\[
    \dot{\theta}_i = \sum_{k=1}^{i}\dot{\beta}_k.
\]
The endpoint velocities are
\begin{align}
    \dot{x}_i
        &=
        \dot{x}
        -\sum_{m=1}^{i-1}\ell_m s_m\dot{\theta}_m
        +\sum_{m=1}^{i-1}c_m\dot{\ell}_m, \nonumber\\
    \dot{y}_i
        &=
        \dot{y}
        +\sum_{m=1}^{i-1}\ell_m c_m\dot{\theta}_m
        +\sum_{m=1}^{i-1}s_m\dot{\ell}_m.
    \label{eq:app_conversion_velocities}
\end{align}
Equivalently, the nonzero entries of $\mathsf{C}$ are
\begin{align}
    C_{x_i,x} &= 1, &
    C_{y_i,y} &= 1, \nonumber\\
    C_{x_i,\beta_k}
        &=
        -\sum_{m=k}^{i-1}\ell_m s_m
        \quad (k<i), &
    C_{y_i,\beta_k}
        &=
        \sum_{m=k}^{i-1}\ell_m c_m
        \quad (k<i), \nonumber\\
    C_{x_i,\ell_m}
        &=
        c_m
        \quad (m<i), &
    C_{y_i,\ell_m}
        &=
        s_m
        \quad (m<i), \nonumber\\
    C_{\theta_i,\beta_k}
        &=
        1
        \quad (k\leq i), &
    C_{\ell_i,\ell_i} &= 1,
    \label{eq:app_conversion_matrix}
\end{align}
with all omitted entries equal to zero. The hydrodynamic matrix in Eq.~\eqref{eq:linear_system} is therefore
\[
    \mathsf{M}=\mathsf{A}\mathsf{C}.
\]
For pinned boundary conditions, the first two rows of $\mathsf{M}$ are replaced by the geometric constraints on $\dot{x}$ and $\dot{y}$. For clamped boundary conditions, the third row is also replaced by the constraint on $\dot{\theta}$.

\section{Cell culture and observation}
\label{app:expr}

{\it Lacrymaria olor} cells were isolated from a water sample collected from the Otogawa River in Okazaki, Aichi, Japan, and maintained 
in Minami-Alps Natural Mineral Water (Suntory) at 25 degrees.
Cultures were fed twice weekly with {\it Cryptomonas} sp. Before experiments, {\it L. olor} cells were transferred to fresh Minami-Alps Natural Mineral Water and starved for approximately 12 h.

For observations of neck extension in immobilized cells, individual {\it L. olor} cells were gently sandwiched between two coverslips (Matsunami) so that movement of the cell body was mechanically restricted while neck movement was retained. For observations of freely swimming cells, the spacing between the coverslips was adjusted to allow cells to move freely. The distance between the coverslips was controlled by adjusting the volume of liquid placed between them.
Bright-field images were acquired using an inverted microscope (ECLIPSE Ti2, Nikon) equipped with a CMOS camera (ORCA-Fusion, Hamamatsu Photonics). Freely swimming cells were observed using a 10x Plan Apo objective lens (NA = 0.45, Nikon), whereas immobilized cells were observed using a 40x Plan Apo objective lens (NA = 0.95, Nikon).

\bibliographystyle{unsrt}
\bibliography{library}

\end{document}